\documentclass[12pt,preprint]{aastex} 
\usepackage{latexsym,times,flushrt}
\usepackage{graphicx}

\def\,{\relax\ifmmode\mskip\thinmuskip\else\thinspace\fi}
\catcode`\@=11
\def\.{.\spacefactor=\@m}

\shorttitle{Detection of Chlorine in Io's Atmosphere}
\shortauthors{Feaga et al.}

\begin{document}

\title{Detection of Atomic Chlorine in Io's Atmosphere with HST/GHRS}

\author{Lori M. Feaga\altaffilmark{1}, Melissa A. McGrath\altaffilmark{2}, 
Paul D. Feldman\altaffilmark{1}, and Darrell F. Strobel\altaffilmark{1}}

\altaffiltext{1}{Department of Physics and Astronomy, Johns Hopkins 
University, 3400 N. Charles Street, Baltimore, MD 21218; 
lanier@pha.jhu.edu, pdf@pha.jhu.edu, strobel@pha.jhu.edu}
\altaffiltext{2}{Space Telescope Science Institute, 3700 San Martin Drive, 
Baltimore, MD 21218; mcgrath@stsci.edu}

\begin{abstract}

We report the detection of atomic chlorine emissions in the atmosphere of 
Io using Hubble Space Telescope observations with the Goddard High Resolution 
Spectrograph (GHRS).  The \ion{Cl}{1} $\lambda$1349 dipole allowed and 
\ion{Cl}{1}] $\lambda$1386 forbidden transition multiplets are detected at a 
signal to noise ratio (SNR) of 6 and 10, respectively, in a combined GHRS 
spectrum acquired from 1994 through 1996.  Oxygen and sulfur emissions are 
simultaneously detected with the chlorine which allows for self-consistent 
abundance ratios of chlorine to these other atmospheric species.  The disk 
averaged ratios are:  Cl/O $=$ 0.017 $\pm$ 0.008, Cl/S $=$ 0.10 $\pm$ 0.05, 
and S/O $=$ 0.18 $\pm$ 0.08.  We also derive a geometric albedo of 
1.0$\,\pm\,$0.4\,$\%$ for Io at 1335\,\AA\ assuming an SO$_2$ atmospheric 
column density of 1\,$\times$10$^{16}$\,cm$^{-2}$.
\end{abstract}

\keywords{planets and satellites: individual (Io)---ultraviolet: solar 
system}

\section{INTRODUCTION}

Io is the most volcanically active body in the solar system.  First viewed by 
Voyager \citep{mor79}, observable volcanic activity revived the 
search for an atmosphere around Io.  After several years of multi-wavelength 
studies \citep{mcg04,lea96,spe96}, it is known that a tenuous atmosphere 
maintained by sublimation and volcanic outgassing and composed primarily of 
SO$_2$ \citep{pea79} and SO \citep{leb96} exists at Io.  This atmosphere is 
denser in the equatorial regions of the satellite and above active volcanic 
plumes \citep{jes04,fel00,mcg00}.  Minor atmospheric species including sulfur 
and oxygen \citep{bal87}, the atomic by-products of SO$_2$, and sodium 
\citep{sch87} have been observed at Io.  Recently, emissions from chlorine 
\citep{re00a}, and NaCl \citep{lel03} have also been detected in the 
atmosphere with inferred abundances lower than those of sulfur and oxygen.  
Extended clouds of potassium have been reported \citep{tra75} suggesting 
potassium local to Io, and \citet{gei04} identified potassium as the most 
probable source of IR emission detected in the equatorial glows, but it has 
not yet been confirmed in the bound atmosphere. 

Since Io's orbit around Jupiter takes it into and out of the densest regions 
of the Io plasma torus, electrons in the torus, co-rotating with Jupiter's 
magnetic field, continually bombard Io's atmosphere, collisionally exciting 
and ionizing atmospheric species. The excited species de-excite by emitting 
photons and produce a rich ultraviolet spectrum.  The ionized material can be 
captured by the magnetic field and populates the torus.  This and other loss 
mechanisms account for a supply of $\sim\,$10$^{30}$ amu per second from Io 
to the jovian magnetosphere.  As a result, elements observed in the 
atmosphere are expected to be components of the torus.  Likewise, species 
detected in the torus should be found in some form in Io's atmosphere.  

\citet{kup00} reported the discovery of Cl$^+$ in the Io plasma torus in the 
near infrared.  From their measurements, the abundance of chlorine ions was 
estimated to be 4\% relative to sulfur ions, indicating a chlorine to total 
ion mixing ratio of 2\%.  For their abundance calculations, they assumed that 
S$^+$, S$^{2+}$, O$^+$, and Cl$^+$ ions constitute 85\% of the torus ion 
density.  In a follow up study by Schneider, Park, \& K\"uppers (2000), 
Cl$^{2+}$ was detected at comparable levels to Cl$^+$, implying a slightly 
larger chlorine to total ion mixing ratio for the torus than first estimated.  
In a more recent study, \citet{fel01} detected both Cl$^+$ and 
Cl$^{2+}$ in a far-UV spectrum of the torus and derived a Cl$^{2+}$ abundance 
of 3\% relative to S$^{2+}$, implying an abundance ratio of 2\% for all 
chlorine ions to all sulfur ions, and a ratio of chlorine to all torus ions 
of at most 1\%.  In 2001, \citet{fel04} acquired better signal to noise ratio 
spectra and confirmed the presence of chlorine ions in the torus, but with a 
30\% lower chlorine to sulfur abundance than previously inferred.  This 
reduction in chlorine abundance observed in data acquired one year later is 
highly suggestive of a variable source of chlorine.  Ionized sodium has also 
been tentatively detected in the torus with a similar estimated abundance of 
1\% \citep{hal94}.

The presence of chlorine ions in the torus has motivated the search for 
chlorine in the atmosphere of Io.  If the atmospheric abundance is similar to 
the $\leq\,$1\% torus abundance, then an atmospheric chlorine column density 
of $\sim\,$1\,$\times$10$^{14}$\,cm$^{-2}$ is implied assuming a nominal 
SO$_{2}$ column density of 1\,$\times$10$^{16}$\,cm$^{-2}$ 
\citep{fel00,mcg00,str01,lel03}.  This implies that neutral chlorine, like 
oxygen and sulfur, should be a detectable component of Io's atmosphere.  
Chlorine is known to be outgassed in terrestrial volcanos and thermochemical 
models show NaCl as the dominant sodium and chlorine bearing gas in Io's 
volcanic gases (Fegley \& Zolotov 2000; Moses, Zolotov, \& Fegley 2002b).  
This prompted a search for NaCl, which was successfully detected by 
\citet{lel03} on both the orbital leading and trailing hemispheres of Io.  
They inferred a disk averaged NaCl abundance of 0.3\% relative to SO$_2$.  
Some of the observed NaCl emission lines are broader than the SO$_2$ emission 
lines, implying locally enhanced NaCl or higher NaCl temperatures.  Since NaCl 
is seen on both hemispheres, may be locally enhanced, and has a short lifetime 
to photolysis, \citet{lel03} inferred that volcanic activity is the most 
likely source of the NaCl.  A variable volcanic source of chlorine could also 
explain the change in chlorine torus abundance reported by \citet{fel04}.

As a result of rapid NaCl photolysis in Io's upper atmosphere, atomic chlorine 
and sodium are more abundant than NaCl \citep{mo02b}.  In an independent study 
of data acquired on two dates by the Hubble Space Telescope (HST) Space 
Telescope Imaging Spectrograph (STIS), \citet{ret02} detected two chlorine 
emission lines in the equatorial spots of Io and calculated a chlorine to 
oxygen abundance of 0.07--1\%.  Here we present unambiguous detections of 
two ultraviolet multiplets of atomic chlorine present in three years of 
archived HST data which provide constraints on the atmospheric chlorine 
abundance.  Although the data show temporal variations over the three 
years of acquisition, this data set gives the most complete disk and 
time averaged abundances for the minor species in Io's atmosphere and aids in 
understanding the nominal atmospheric composition.  We first describe the 
observations and data reduction, then present the analysis of the chlorine 
detection, giving estimates of the relative chlorine abundances at Io, and 
finally, discuss the implications of our findings. 

\section{OBSERVATIONS AND DATA REDUCTION}

We utilize an extensive set of Goddard High Resolution Spectrograph (GHRS) 
archival spectra acquired during 1994--1996 with the 
1.74$''\,\times$\,1.74$''$ large science aperture and G140L grating covering 
the spectral range 1250--1500\,\AA\.  Although preliminary analyses of these 
data have been presented \citep{bal96,bal97,re00a}, none have been previously 
published.  The dataset consists of 25 exposures varying in length from 6 to 
26 minutes (see Table 1 for details).  We concentrate in this paper only on 
the 1330--1420\,\AA\, portion of the spectrum because it contains previously 
unidentified features as well as known oxygen and sulfur emissions.  

The data are reduced in the following manner.  A flux spectrum is created by 
co-adding 5.5 hours of data to maximize the signal to noise ratio (SNR).  
Spectra from the individual dates are presented in Figure 1.  Large temporal 
variations can be seen in the data and are most likely to be caused by 
differences in the spatial distribution and abundance of the emitting species 
in Io's atmosphere and the variable emission morphology \citep{roe99,re00b} 
driven by the interaction between Io and its changing electromagnetic 
environment.  Due to these effects, the line profiles are not necessarily 
symmetric or well centered at the laboratory wavelength of the transition, 
which can result in systematic wavelength shifts.  Accordingly, when 
co-adding, the data are weighted by exposure time and aligned in wavelength 
using the strong \ion{O}{1}] $\lambda$1356 emission feature in each spectrum.  
The errors are determined from the observed count rate using Poisson 
statistics and propagated through the rest of the reduction steps.  A 
background subtraction is done to correct for an erroneous slope introduced 
during the original pipeline calibration in which only the four corner diodes 
(which have virtually no counts) were used to determine the detector 
background.  The resulting spectrum is smoothed by 3 bins.  A nominal full 
width at half maximum (FWHM) of 3.2\,\AA\, for the observed features is 
determined by $\chi^2$ fitting Gaussian profiles to the strongest and least 
blended emission multiplet in the spectrum, \ion{O}{1}] $\lambda$1356.  We 
assume this value is constant across the spectrum and use it in subsequent 
analysis.  The reduced, co-added flux spectrum is shown in Figure 2.  
Previously identified oxygen and sulfur emissions from the Io system are 
noted \citep{bal87,moo91,dur95}.  Previously unidentified emissions are 
indicated by the X's in the figure.

The first step in our analysis is to identify the lines present in the data 
by creating synthetic spectra for the 1330--1420\,\AA\, region using only the 
known oxygen and sulfur transitions listed in Table 2.  The synthetic spectra 
are comprised of Gaussians with FWHM of 3.2\,\AA\, at each of the central 
wavelengths of the oxygen and sulfur transitions.  For identification 
purposes only, the relative amplitudes of the Gaussians are constrained to 
satisfy the optically thin line ratios (Table 2).  Next, best fit spectra are 
determined for the confirmed sulfur and oxygen multiplets in the data using 
SPECFIT, an IRAF $\chi^2$ minimization spectral fitting routine 
\citep{kri94}.  The FWHM and number of Gaussians to be fit in the spectral 
region are fixed and supplied to the program while the amplitude of each line 
is a free parameter, and the central wavelength of each line is restricted to 
a range of $\pm$ 1.4\,\AA\, from the laboratory wavelength.  This wavelength 
constraint is used because of the instrumental wavelength uncertainty of 
0.7\,\AA\, and the variable observational wavelength shift mentioned 
above.  Line fluxes extracted using the best fit Gaussian profiles along with 
the best fit central wavelengths are summarized in Table 2.  Our best fit to 
the \ion{O}{1}] $\lambda$1356 doublet is shown in Figure 3, with the 
transitions displaying a ratio of 3.3 $\pm$ 0.4 to 1, consistent with the 
expected optically thin ratio.  

The \ion{S}{1} $\lambda$1389 and \ion{S}{1} $\lambda$1405 spectral fits are 
shown in Figure 4a.  There are six components in the \ion{S}{1} $\lambda$1389 
multiplet and three components in the \ion{S}{1} $\lambda$1405 multiplet 
(Table 2).  Atomic data for the UV sulfur transitions are limited, especially 
for the \ion{S}{1} $\lambda$1389 multiplet which has only two published 
oscillator strength values for the individual transitions, one theoretical 
\citep{var61} and one experimental \citep{mul68}, differing by several orders 
of magnitude.  The 3s\,3p$^{5}$\,{}$^{3}$P\arcdeg\, state of this multiplet 
is highly mixed, difficult to correctly represent in theoretical 
calculations, and in Io's environment is excited by low-temperature, 
near-threshold electrons.  Because of this, neither the experimental nor the 
theoretical data are ideal. Although both give similar line ratios, 
\citet{mor03} reports the experimental data \citep{mul68} which we also 
utilize.  Our observed relative strengths significantly disagree with the 
expected optically thin values.  Specifically, transitions 
with an upper level of $J$=2, 1388\,\AA\, and 1396\,\AA, are 2--3 times larger 
than expected.  However, our values are reasonably consistent with those 
observed by \citet{ret02}.  What looks to be a large discrepancy between the 
two data sets in the relative strengths of the 1388/1389 lines is a result of 
blending of these two transitions.  If analyzed as a pair rather than 
individual lines, comparable values of $\sim\,$5 $\pm$ 2 for (1388+1389)/1381 
emerge.  We do not attribute the disagreement between observed data from Io 
and published atomic data for \ion{S}{1} $\lambda$1389 to optical depth 
effects since the multiplet oscillator strength is 2--3 orders of magnitude 
smaller than those of other allowed sulfur multiplets and is comparable in 
magnitude to the forbidden lines, which do not experience self-absorption.  
In addition, we find that the strongest lines in the data, which are 
expected to be strongest in the optically thin case and hence would 
be most affected by optical depth, are the transitions for which the ratios 
are inconsistently large, in the opposite sense of optical thickness effects.

The relative strengths of the \ion{S}{1} $\lambda$1405 multiplet also differ 
from the expected optically thin values, which we attribute to optical 
thickness effects and spectral contamination by the Io plasma torus.  The 
weakest line, 1413\,\AA, is marginally detected while the remaining lines, 
1401\,\AA\, and 1409\,\AA, are comparable in strength, corresponding to the 
redistribution of line flux from the strongest line, 1401\,\AA, to the longer 
wavelength line, 1409\,\AA.  The oscillator strength, optical thickness 
effects, and configuration of this multiplet resemble that of the 
\ion{S}{1} $\lambda$1814 allowed multiplet in Io's atmosphere which has also 
been interpreted to be optically thick (Feaga, McGrath, \& Feldman 2002; 
McGrath et al. 2000; Ballester et al. 1987).  The ratio of 1409\,\AA\, to 
1413\,\AA, however, is much higher than expected even in the optically thick 
case where 1409\,\AA\, is preferentially pumped by 1401\,\AA\, and is the 
result of \ion{S}{4} torus emission blending with the \ion{S}{1} $\lambda$1405 
multiplet, as explained below.  

We investigate a feature longward of the \ion{S}{1} $\lambda$1405 multiplet, 
marked with a Y in Figure 2, which is detected at a SNR of 5 and is 
coincident in wavelength with a \ion{S}{4} transition at 1416.9\,\AA\, to 
determine if the \ion{S}{4} multiplet, composed of 5 transitions from 
1398--1424\,\AA, contributes to the inconsistent line ratios discussed above 
for the \ion{S}{1} $\lambda$1405 multiplet.  Using the spectral fitting 
routine, and allowing the FWHM of the \ion{S}{4} multiplet to differ from the 
emissions from Io, the weak 1398\,\AA\, line is not detected.  The 1424\,\AA\, 
line is coincident with a much stronger \ion{S}{1} $\lambda$1429 multiplet 
and we do not attempt to disentangle its expected minimal contribution.  
Fits to the remaining two lines, 1404\,\AA\, and 1406\,\AA, provide a better 
overall fit to the data, although the error bars on the 1404\,\AA\, line 
are large and contribute to the uncertainty of the \ion{S}{1} $\lambda$1405 
fluxes.  The 1406\,\AA\, and 1417\,\AA\, lines are expected to be the 
strongest lines of \ion{S}{4}] $\lambda$1406 (Figure 4a).  As detected, they 
have relative line strengths consistent with those found using atomic data 
from \citet{tay99,tay00} and observed in the Io torus \citep{moo81,moo91}.  We 
tentatively identify these emissions as \ion{S}{4} originating from the 
Io plasma torus since the GHRS field of view is larger than the disk of Io 
and the line of sight from HST to Io intersects the torus.

Additional emission, not associated with the oxygen and sulfur features, is 
obvious in Figures 4 and 5 flanking the \ion{O}{1}] $\lambda$1356 doublet and 
the \ion{S}{1} $\lambda$1389 multiplets.  Emission near 1335\,\AA\, and 
1379\,\AA\, is seen in spectra from all 8 dates of observation (Figure 1).  
In the co-added flux spectrum, the 1379\,\AA\, emission is detected at a SNR 
of 10 (Figure 4).  The emission features at 1335\,\AA, 1347\,\AA, and 
1363\,\AA\, (Figure 5) are detected at a SNR of $\sim\,$6.

Two multiplets of chlorine occur in the 1330--1400\,\AA\ region, a dipole 
allowed transition, \ion{Cl}{1} $\lambda$1349, consisting of four lines and 
\ion{Cl}{1}] $\lambda$1386, a forbidden transition consisting of five lines 
(Table 2).  Synthetic profiles of these multiplets using optically thin line 
ratios are shown in Figures 4b and 5b.  Using the same $\chi^2$ minimization 
method described above in which the amplitude is a free parameter, wavelength 
is constrained, and FWHM is fixed, fits are made to these previously 
unidentified features and are presented in Figure 4c and 5c.  Using data 
acquired by STIS with higher spectral resolution but lower SNR, \citet{ret02} 
presents the detection of one chlorine line per multiplet, 1347.2\,\AA\, 
and 1379.5\,\AA. 

In the \ion{Cl}{1}] $\lambda$1386 multiplet, lines other than 1379.5\,\AA, 
which is the strongest line of the multiplet accounting for more than 75\% of 
the multiplet's total strength, are either too weak or blended with the 
sulfur to unambiguously identify.  The component at 1373.1\,\AA\, is not 
detected, consistent with its very small strength compared to the 
1379.5\,\AA\, line.  We neglect all but the 1379.5\,\AA\, line in the 
subsequent flux analysis.

The flux ratio of 1347.2\,\AA\, to 1363.5\,\AA\, in the 
\ion{Cl}{1} $\lambda$1349 multiplet is almost a factor of 4 lower than 
expected in the optically thin case, indicating the redistribution of flux 
between the two lines.  Because of this, we infer that the multiplet is 
optically thick.  The flux ratio of 1335.7/1351.7 is 2.6, more than a factor 
of 5 higher than the expected optically thin value for the two 
\ion{Cl}{1} $\lambda$1349 lines at these wavelengths.  As discussed further 
below, we attribute the majority of the integrated flux for the 1335\,\AA\, 
feature to a factor other than optical depth effects.  

The 1335.7\,\AA\, feature of the \ion{Cl}{1} $\lambda$1349 multiplet is not 
well fit with one Gaussian of 3.2\,\AA\, FWHM.  The best fit for a single 
Gaussian at 1335.7\,\AA\, results in a line with a FWHM of 1.8\,\AA.  
Considering the unusually high flux and the smaller FWHM of the 1335.7\,\AA\, 
feature, we attribute the majority of the integrated flux at 1335.7\,\AA\, 
not to chlorine but to the strong \ion{C}{2} $\lambda$1335 emission multiplet, 
with a component at 1335.7\,\AA, in the solar spectrum reflected from Io and 
detected previously from both Europa and Ganymede \citep{hal95,hal98}.  We 
model the reflection of the \ion{C}{2} $\lambda$1335 multiplet from Io's disk 
and determine that \ion{C}{2} has a smaller FWHM than Iogenic emissions for 
which the FWHM are broadened by the geometry of the equatorial spots and limb 
glow morphology.  To correctly account for the solar carbon and Iogenic 
chlorine contribution to 1335\,\AA, we re-fit the feature with three 
Gaussians, two representing the 1334.5\,\AA\, and 1335.7\,\AA\, lines of 
\ion{C}{2} with FWHM of 1.4\,\AA\, and one chlorine line with a FWHM of 
3.2\,\AA\, (Figure 5c).  These fluxes are listed in Table 2.

This is the first reported detection of \ion{C}{2} $\lambda$1335 from Io, 
providing a unique measurement of the UV geometric albedo of the satellite at 
this wavelength.  Following \citet{fel00}, Io's albedo is calculated using a 
weighted average of solar spectra from the UARS SOLSTICE experiment 
\citep{woo96} acquired on the dates of HST/GHRS Io observation.  All 
observations occurred during periods of minimal solar activity.  With no 
SO$_2$ absorption, the best fit to the data in Figure 5c gives a disk 
averaged geometric albedo of 0.83$\pm$0.35\%.  However, the albedo calculation 
at 1335\,\AA\, is sensitive to the SO$_2$ column density.  An SO$_2$ column 
density of $\sim\,$10$^{16}$\,cm$^{-2}$ would increase the derived albedo to 
1.0\%.  \citet{fel00} estimate albedos of 1.5\% for HST/STIS data acquired in 
1997 and 1.9\% for 1998 by fitting the continuum between 1520\,\AA\ and 
1700\,\AA\.  In this wavelength interval, the albedo calculation is much less 
sensitive to the SO$_2$ column since the absorption cross section of SO$_2$ 
is half that at 1335\,\AA\.  Including a latitudinally varying surface albedo 
and a modeled SO$_2$ column density, which when disk averaged is 
$\sim\,$1$\times$10$^{16}$\,cm$^{-2}$, \citet{str01} derive albedos ranging 
from 2.2--3.1\% to match the mid-latitude Lyman-$\alpha$ measurements in the 
1998 HST/STIS data.  An SO$_2$ column density of 
3.6--8.0$\times$10$^{16}$\,cm$^{-2}$ would easily bring our albedo value into 
agreement with those of \citet{fel00} and \citet{str01} which is consistent 
with the known properties of Io's SO$_2$ atmosphere \citep{mcg04}.
Differing values of the geometric albedo estimated by \citet{fel00} and 
\citet{str01} for the same data using similar SO$_2$ column densities may 
suggest a wavelength dependence of the UV albedo rather than measurable 
variations in the SO$_2$ atmosphere.  

We have also considered the contribution of solar resonant scattering to the 
\ion{Cl}{1} $\lambda$1349 multiplet.  As stated previously, 
\ion{Cl}{1} 1335.7\,\AA\, is coincident in wavelength with a strong solar 
\ion{C}{2} transition.  Since the \ion{Cl}{1} 1335.7\,\AA\, and 1351.7\,\AA\, 
lines share the same upper level, it is possible for 1351.7\,\AA\, to be 
pumped by the strong solar carbon line at 1335.7\,\AA.  Including a curve of 
growth to account for self-absorption in our data where the 
\ion{Cl}{1} $\lambda$1349 multiplet is optically thick, the fluorescence 
saturates at 0.02$\,\times$\,10$^{-4}\,$photons\,cm$^{-2}\,$s$^{-1}$.  We 
conclude that the induced fluorescence of 1351.7\,\AA\, is negligible.

\section{ANALYSIS}

Since our data set lacks spatial information, we rely on \citet{ret02} 
for information about the spatial distribution of the chlorine emission at 
Io.  He finds that the emitting chlorine has similar morphology to that of 
atomic oxygen and sulfur, and is therefore assumed to be located at higher 
altitudes than the bulk of the SO$_2$ and to be excited by the same electron 
impact mechanism as oxygen and sulfur.  This assumption allows us to give 
concurrent estimates of the relative abundances of these species from the 
same spectrum.  For a transition excited by electron impact, the total 
measured brightness for a multiplet along the line of sight ($B_{ij}$) is 
related to the density of the emitting species ($n_s$), the electron density 
($n_e$), and the temperature-dependent electron excitation rate coefficient 
($Q_{ij}$) of the multiplet:
$$
{B_{ij}(T_e)} = \int { n_s n_e Q_{ij}(T_e) dl}.
\eqno(1)
$$
In calculating ratios of species detected along the same line of sight, the 
dependence on the electron density and line of sight conditions is removed 
and an approximate proportionality is established as a function of electron 
temperature ($T_e$).  Since the electron impact excitation rate coefficient 
is a function of electron temperature, we choose a nominal value 5 eV for 
$T_e$.  Incorporating the \ion{Cl}{1} $\lambda$1349 and 
\ion{O}{1}] $\lambda$1356 cross sections from \citet{gan88} and D. F. Strobel 
(2002, private communication), respectively, \citet{ret02} made detailed 
calculations of the rate coefficients which are adopted here (Table 3).  
Using recently published atomic data \citep{zat02}, the electron impact 
excitation rates for several sulfur multiplets are calculated and listed in 
Table 3.  A liberal error estimate of 30\% for the rate coefficients is 
propagated through the relative abundance calculations.

To achieve the best estimate of the Cl/O, Cl/S, and S/O abundance ratios, 
total measured fluxes of \ion{Cl}{1} $\lambda$1349, 
\ion{O}{1}] $\lambda$1356, \ion{S}{1} $\lambda$1429, 
and \ion{S}{1} $\lambda$1479 are selected for the calculation.  
\ion{Cl}{1} $\lambda$1349 is chosen since it is the least 
blended chlorine multiplet and the only chorine multiplet in our data for 
which reliable atomic data have been published.  Two allowed sulfur multiplets 
are selected in lieu of forbidden sulfur multiplets and the contaminated 
\ion{S}{1} $\lambda$1405 multiplet.  \ion{S}{1} $\lambda$1429 and 
\ion{S}{1} $\lambda$1479 are dominant sulfur transitions in the spectral 
range of the data with high SNR and better atomic data than the other sulfur 
multiplets.  Following \citet{fea02}, we assume a 40\% : 60\% contribution to 
the total 1479\,\AA\, flux of the individual forbidden and allowed 
multiplets, respectively, as they are not resolved in the GHRS data.  With 
these fluxes and the corresponding electron rate coefficients from Table 3, 
we estimate and present in Table 4 disk averaged abundance ratios.  Because 
of the large errors for the abundance ratios due in part to the uncertainty 
in the electron impact rate coefficients, the results are insensitive to 
$T_e$ for a range of 3--7 eV.  

To quantify the contribution of the equatorial spot emission to the total 
emission from Io, we present the STIS equatorial spot to GHRS disk averaged 
emission flux ratio for \ion{Cl}{1} $\lambda$1349 and compare it to the spot 
to disk ratio of \ion{O}{1}] $\lambda$1356 in STIS data alone.  The 
equatorial spot flux listed in Table 2 is the average flux for a single spot 
extracted using a 0.049 (arcsec)$^{2}$ solid angle \citep{ret02} and the disk 
flux is calculated from the emission contained in the 1.74$''$ 
square GHRS aperture.  The equatorial spots are 2--5\% of the total disk 
emission for chlorine.  For comparison, an oxygen spot to disk ratio of 
2--5\% is estimated using the data from \citet{ret02}.  The chlorine and 
oxygen spot to disk values are similar.

\section{DISCUSSION}

Utilizing the definition of optical depth, $\tau\,=\,{\cal N}\sigma$, where 
${\cal N}$ is the column density and $\sigma$ is the absorption cross 
section, and the determination of the \ion{Cl}{1} $\lambda$1349 multiplet 
being optically thick in our data ($\tau>\,$1), a lower bound to the disk 
averaged chlorine column density of 2.3$\times$10$^{12}$\,cm$^{-2}$ is 
implied.  The chlorine absorption cross section is calculated with the 
oscillator strength from \citet{bie94} at a temperature of 1000 K.  In the 
same manner, the \ion{S}{1} $\lambda$1405 multiplet appears to be optically 
thick in our data giving a lower bound to the disk averaged sulfur column 
density of 2.4$\times$10$^{13}$\,cm$^{-2}$ using an oscillator strength from 
\citet{tay98}.  This result is consistent with the disk averaged sulfur 
column density lower bound estimated by \citet{fea02} for the 
\ion{S}{1} $\lambda$1814 multiplet in IUE and FOS data.  This independent 
estimate of the sulfur column density lends credence to the range of 
sulfur column densities found by \citet{fea02}.

The S/O results presented here are in agreement with the \citet{wol01} 
analysis of HST/STIS data.  They find a disk averaged S/O brightness ratio of 
1 : 0.81 when comparing the combined \ion{S}{1} $\lambda$1479 multiplets 
to the \ion{O}{1}] $\lambda$1356 doublet.  Applying the 40\% : 60\% 
contribution of the forbidden and allowed multiplets \citep{fea02}, 
respectively, to the \citet{wol01} analysis and utilizing the rate 
coefficients from Table 3, their brightness ratio translates to an S/O 
abundance ratio of 0.10 $\pm$ 0.06.  
  
Combining the S/SO$_2$ ratio of 3--7$\times$10$^{-3}$ measured at three 
spatially resolved locations by \citet{mcg00} with our results, an estimate 
of the Cl/SO$_2$ ratio can be made, which leads to an approximate Cl/SO$_2$ 
abundance ratio of 3.0--7.0$\times$10$^{-4}$ (Table 4).  In addition, 
combining the O/SO$_2$ ratio of 0.05 inferred with Cassini data \citep{gei04} 
with our results gives a Cl/SO$_2$ value of 8.5$\times$10$^{-4}$.  Recently, 
\citet{lel03} detected a comparable abundance of NaCl with a disk averaged 
NaCl/SO$_2$ ratio of 7$\times$10$^{-4}$--0.02 if NaCl is more localized than 
SO$_2$, with a preferred value of 3.5$\times$10$^{-3}$ (Table 4).  

In assessing escape rates of sodium, chlorine, and NaCl from Io, \citet{mo02b} 
establish that chlorine and sodium have escape fluxes 10--20 times larger 
than NaCl.  Our estimated atmospheric chlorine abundance and the 
NaCl abundance of \citet{lel03} are both 1--2 orders of magnitude lower than 
the 1\% torus chlorine mixing ratio \citep{fel01} indicating that there 
exists an intricate relationship between the transport of volcanic gases to 
the upper atmosphere and escape to the torus.  There may also be a higher 
concentration of chlorine compared to sulfur and oxygen near volcanic 
outflows which are at lower altitudes than we probe and would boost 
the overall atmospheric chlorine abundance measured in our data.
In the Pele-type volcanic atmospheres modeled by \citet{mo02b} 
which include chlorine, sodium, and potassium, the initial chlorine abundance 
is assumed from the high chlorine torus abundance of \citet{sch00}.  They do 
not check the sensitivity of their models to lower chlorine abundances, 
so we do not compare our results for chlorine.  We do however compare the S/O 
abundance ratios.  \citet{mo02a} determine an equilibrium ratio of 10 for S/O, 
$\sim\,$50 times larger than our data shows.  This suggests that either the 
initial oxygen abundances in the models are too low or the disk averaged 
atmosphere we detect at Io is dominated by Prometheus-type (SO$_2$ dominated) 
outflows rather than Pele-type (sulfur dominated).  \citet{mo02a} do not 
check the sensitivity of their models to the initial S/O abundance.

Lastly, our Cl/S ratio adds plausibility to the identification of solid 
Cl$_2$SO$_2$ or ClSO$_2$ in NIMS/Galileo spectra near the volcanic center of 
Marduk as suggested by \citet{sch03}.  Unidentified absorption features are 
present in the infrared spectra and based on their spectroscopic arguments, 
Cl$_2$SO$_2$ is the favorite candidate with ClSO$_2$ a potential alternative.  
\citet{sch03} estimate that for abundant Cl$_2$SO$_2$ formation, the gaseous 
[Cl-(Na+K)]/S ratio in a volcanic plume needs to be larger than 0.015, and for 
ClSO$_2$, the ratio must be even larger.  When there is an excess of halogens 
as compared to alkalis, Cl/(Na+K) $>$ 1, sulfur chlorides and oxychlorides 
are formed with the surplus chlorine.  Since volcanic outgassing is 
assumed to be the only source of chlorine in Io's atmosphere, plumes should 
have at least a comparable, most likely higher, abundance of chlorine 
relative to the disk averaged atmosphere.  For comparable Cl and NaCl 
disk averaged abundances, consistent with our data and that of \citet{lel03}, 
an estimated Cl/Na ratio of $\sim\,$10 is shown in Figure 3d of 
\citet{feg00}, and for our Cl/S ratio of 0.1, a Cl/Na ratio $\geq\,$2 can be 
estimated from Figure 5 of \citet{feg00}.  These ratios imply that gaseous 
chlorine is at least twice as high in abundance as sodium.  Therefore, our 
disk averaged Cl/S ratio is consistent with the requirements of \citet{sch03} 
for the presence of Cl$_2$SO$_2$ or ClSO$_2$.

\acknowledgments

The authors would like to thank Kurt Retherford and Warren Moos for useful 
discussion of the chlorine detections in the STIS data.  Support 
for this work was provided by NASA through grant number HST-AR-09211.01-A 
from the Space Telescope Science Institute, which is operated by the 
Association of Universities for Research in Astronomy, Incorporated, under 
NASA contract NAS5-26555.


\clearpage
\begin{deluxetable}{llrrcc}
\tabletypesize{\footnotesize}
\tablecaption{Table of Observations \label{tbl-1}}
\tablewidth{0pt}
\tablecolumns{6}
\tablehead{
\colhead{Obs Name} & \colhead{Date} & \colhead{Time} &
\colhead{Exposure} & \colhead{Io OLG\tablenotemark{a}} & \colhead{Io SYS III}\\
\colhead{} & \colhead{} & \colhead{UT} &\colhead{(s)} & 
\colhead{($\arcdeg$)} & \colhead{($\arcdeg$)}\\
}\startdata

z2eh0506 & 1994 Jun 06 & 16:48:41 &  620.0 & \,\,98 & 140 \\
z2eh0508 & 1994 Jun 06 & 18:06:11 & 1580.0 & 109 & 176 \\
z2eh0306 & 1994 Jun 14 & 15:57:23 & 1580.0 & 279 & \,\,52 \\
z2eh0308 & 1994 Jun 14 & 17:34:05 &  620.0 & 293 & \,\,97 \\
z2eh0306 & 1994 Jun 16 & 09:48:23 & 1580.0 & 274 & 135 \\
z2eh0408 & 1994 Jun 16 & 11:24:37 &  600.0 & 287 & 176 \\
z2gd0e02 & 1994 Jul 15 & 05:24:16 & 1555.5 & \,\,17 & 276 \\
z2v9a102 & 1995 Sep 21 & 20:03:23 &  972.8 & \,\,18 & \,\,73 \\
z2v9a104 & 1995 Sep 21 & 20:26:59 &  942.4 & \,\,21 & \,\,84 \\
z2v9a106 & 1995 Sep 21 & 21:33:59 &  972.8 & \,\,30 & 116 \\
z2v9a108 & 1995 Sep 21 & 21:58:23 &  790.4 & \,\,34 & 126 \\
z2v90308 & 1996 Mar 23 & 19:20:52 &  972.8 & 338 & \,\,31 \\
z2v9030a & 1996 Mar 23 & 19:44:34 &  942.4 & 341 & \,\,43 \\
z2v9020a & 1996 Jun 26 & 04:26:53 &  851.2 & 106 & 350 \\
z2v9020e & 1996 Jun 26 & 06:03:22 &  851.2 & 119 & \,\,35 \\
z2v9020h & 1996 Jun 26 & 07:21:40 &  699.2 & 130 & \,\,72 \\
z2v9020i & 1996 Jun 26 & 07:35:53 &  364.8 & 132 & \,\,77 \\
z2v9020k & 1996 Jun 26 & 08:49:34 &  760.0 & 143 & 112 \\
z2v9020m & 1996 Jun 26 & 09:10:10 &  790.4 & 146 & 121 \\
z2v9020o & 1996 Jun 26 & 10:26:10 &  851.2 & 156 & 156 \\
z2v9020q & 1996 Jun 26 & 10:48:22 &  851.2 & 160 & 167 \\
z3eta102 & 1996 Sep 04 & 11:09:10 &  972.8 & \,\,12 & \,\,67 \\
z3eta104 & 1996 Sep 04 & 11:32:59 & 1033.6 & \,\,16 & \,\,79 \\
z3eta106 & 1996 Sep 04 & 12:45:10 &  668.8 & \,\,26 & 112 \\
z3eta109 & 1996 Sep 04 & 13:13:16 &  832.0 & \,\,30 & 125 \\

\enddata
\tablenotetext{a}{Orbital Longitude of Io.}
\end{deluxetable}

\clearpage

\begin{deluxetable}{llccllccc}
\tabletypesize{\footnotesize}
\tablecaption{Atomic Transitions and Measured Fluxes \label{tbl-2}}
\tablewidth{0pt}
\tablecolumns{9}
\tablehead{
\colhead{Species} &\colhead{Configuration} & \colhead{$J_u$} & 
\colhead{$J_l$} & \colhead{$\lambda_{lab}$} & \colhead{$\lambda_{fit}$} & 
\colhead{Relative} & \colhead{GHRS Flux\tablenotemark{b}} & 
\colhead{STIS Flux\tablenotemark{c}} \\
\colhead{} & \colhead{} & \colhead{} & \colhead{} & \colhead{(\AA)} & 
\colhead{(\AA)} & \colhead{strength\tablenotemark{a}} & 
\multicolumn{2}{c}{(10$^{-4}\,$photons\,\,cm$^{-2}\,$s$^{-1}$)} \\ 
}\startdata

C II & 2s\,2p$^{2}$\,{}$^{2}$D$\,\to$\,2s$^{2}$\,({}$^{1}$S)\,2p\,{}$^{2}$P\arcdeg & 3/2 & 1/2 & 1334.53 & 1334.30 & 5.0 & 0.08 $\pm$ 0.07 & \\
 &  &  3/2 & 3/2 & 1335.66 &  & 1.0 & & \\
 &  &  5/2 & 3/2 & 1335.71 & 1335.40 & 9.0 & 0.16 $\pm$ 0.07 & \\
\\
\tableline
\\
Cl I & 3s$^{2}$\,3p$^{4}$\,({}$^{3}$P)\,4s\,{}$^{2}$P$\,\to$\,3s$^{2}$\,3p$^{5}$\,{}$^{2}$P\arcdeg & 1/2 & 3/2 & 1335.73 & 1336.80 & 1.1 & 0.13 $\pm$ 0.06 & \\
 &  & 3/2 & 3/2 & 1347.24 & 1348.13 & 5.5 & 0.34 $\pm$ 0.08 & 0.012 $\pm$ 0.003 \\
 &  & 1/2 & 1/2 & 1351.66 & 1351.52 & 2.2 & 0.14 $\pm$ 0.08 & 0.006 $\pm$ 0.004 \\
 &  & 3/2 & 1/2 & 1363.45 & 1363.35 & 1.0 & 0.23 $\pm$ 0.04 & \\ 
\\
\tableline
\\
O I & 2s$^{2}$\,2p$^{3}$\,3s\,{}$^{5}$S\arcdeg$\to$\,2s$^{2}$\,2p$^{4}\,{}^{3}$P & 2 & 2 & 1355.60 & 1355.56 & 3.1 & 7.71 $\pm$ 0.34 & 0.830 $\pm$ 0.020 \\
 &  & 2 & 1 & 1358.51 & 1358.34 & 1.0 & 2.31 $\pm$ 0.28 & 0.230 $\pm$ 0.010 \\
\\
\tableline
\\
Cl I & 3s$^{2}$\,3p$^{4}$\,({}$^{3}$P)\,4s\,{}$^{4}$P$\,\to$\,3s$^{2}$\,3p$^{5}\,{}^{2}$P\arcdeg & 1/2 & 3/2 & 1373.12 & & 1.0 & & \\
 &  & 3/2 & 3/2 & 1379.53 & 1380.14 & 58.8 & 0.67 $\pm$ 0.22\tablenotemark{d} & 0.036 $\pm$ 0.006 \\
 &  & 5/2 & 3/2 & 1389.69 & & 2.8 & &  \\
 &  & 1/2 & 1/2 & 1389.96 & & 7.3 & &  \\
 &  & 3/2 & 1/2 & 1396.53 & & 7.1 & &  \\
\\
\tableline
\\
S I & 3s\,3p$^{5}$\,{}$^{3}$P\arcdeg$\to$\,3s$^{2}$\,3p$^{4}\,{}^{3}$P & 1 & 2 & 1381.55 & 1382.83 & 1.0 & 0.50 $\pm$ 0.16 & 0.042 $\pm$ 0.006 \\
 &  & 0 & 1 & 1385.51 & 1386.34 & 0.8 & 0.45 $\pm$ 0.10 & 0.046 $\pm$ 0.006\\
 &  & 2 & 2 & 1388.44 & 1388.79 & 1.7 & 2.00 $\pm$ 0.35 & 0.130 $\pm$ 0.007\tablenotemark{e}\\
 &  & 1 & 1 & 1389.15 & 1390.24 & 0.3 & 0.65 $\pm$ 0.19 & 0.086 $\pm$ 0.006\tablenotemark{e}\\
 &  & 1 & 0 & 1392.59 & 1393.60 & 1.0 & 1.00 $\pm$ 0.22 & 0.051 $\pm$ 0.006\\
 &  & 2 & 1 & 1396.11 & 1396.56 & 1.8 & 1.42 $\pm$ 0.23 & 0.127 $\pm$ 0.008\tablenotemark{e}\\
\\
\tableline
\\
S IV & 3s\,3p$^{2}$\,$^{4}$P\arcdeg$\to$\,3s$^{2}$\,3p\,${}^{2}$P\arcdeg & 3/2 & 1/2 & 1398.04 & & 0.2 & & \\
 &  & 1/2 & 1/2 & 1404.81 & 1404.80 & 1.0 & 0.06 $\pm$ 0.18 & \\
 &  & 5/2 & 3/2 & 1406.02 & 1406.80 & 3.5 & 0.18 $\pm$ 0.10 & \\
 &  & 3/2 & 3/2 & 1416.89 & 1416.75 & 3.0 & 0.22 $\pm$ 0.01 & \\
 &  & 1/2 & 3/2 & 1423.84 &  & 0.8 &  & \\
\\
\tableline
\\
S I & 3s$^{2}$\,3p$^{3}$\,5s\,{}$^{3}$S\arcdeg$\to$\,3s$^{2}$\,3p$^{4}\,{}^{3}$P & 1 & 2 & 1401.51 & 1402.20 & 5.3 & 0.57 $\pm$ 0.16 & \\
 &  & 1 & 1 & 1409.34 & 1408.90 & 3.1 & 0.54 $\pm$ 0.08 & \\
 &  & 1 & 0 & 1412.87 & 1412.20 & 1.0 & 0.10 $\pm$ 0.08 & \\
\\
\tableline
\\
S I \tablenotemark{f} & 3s$^{2}$\,3p$^{3}$\,3d\,{}$^{3}$D\arcdeg$\to$\,3s$^{2}$\,3p$^{4}\,{}^{3}$P & & & 1429.11 & & & 5.78 $\pm$ 0.22 & \\
\\
\tableline
\\
S I \tablenotemark{f} & 3s$^{2}$\,3p$^{3}$\,3d\,${}^{5}$D\arcdeg$\to$\,3s$^{2}$\,3p$^{4}\,{}^{3}$P & & & 1477.31 & & & 6.69 $\pm$ 0.27 \tablenotemark{g} & \\
\\
\tableline
\\
S I \tablenotemark{f} & 3s$^{2}$\,3p$^{3}$\,4s$'\,{}^{3}$D\arcdeg$\to$\,3s$^{2}$\,3p$^{4}\,{}^{3}$P & & & 1478.50 & & & 10.03 $\pm$ 0.40 \tablenotemark{g} & \\

\enddata
\tablenotetext{a}{Relative line strengths computed for optically thin case 
from atomic data in \citet{mor03} for \ion{O}{1}, \ion{S}{1}, and \ion{C}{2}; 
\citet{bie94} for \ion{Cl}{1}; and \citet{tay99,tay00} for \ion{S}{4}.}
\tablenotetext{b}{HST/GHRS disk averaged data presented in this paper.}
\tablenotetext{c}{HST/STIS equatorial spot data presented in \citet{ret02}.}
\tablenotetext{d}{Flux estimate available for 1379.53\,\AA\, line only.}
\tablenotetext{e}{\citet{ret02} reports these as blends.}
\tablenotetext{f}{Multiplet values only.}
\tablenotetext{g}{Fluxes estimated from blended data assuming 40\% : 60\% 
forbidden to allowed contribution.}
\end{deluxetable}

\clearpage

\begin{deluxetable}{ll}
\tabletypesize{\footnotesize}
\tablecaption{Electron Impact Excitation Rate Coefficients \label{tbl-3}}
\tablewidth{0pt}
\tablecolumns{2}
\tablehead{
\colhead{Species} &\colhead{Rate Coefficient\tablenotemark{a}} \\
\colhead{} & \colhead{(cm$^{3}$ s$^{-1}$)} \\ 
}\startdata

\ion{Cl}{1} $\lambda$1349 & 1.6$\times$10$^{-9}$  \\ 
\ion{O}{1}] $\lambda$1356 & 3.2$\times$10$^{-10}$ \\
\ion{S}{1} $\lambda$1429  & 1.1$\times$10$^{-9}$  \\
\ion{S}{1} $\lambda$1479  & 1.8$\times$10$^{-9}$  \\

\enddata
\tablenotetext{a}{Electron temperature of 5 eV used in the rate coefficient 
calculation.}
\end{deluxetable}

\clearpage

\begin{deluxetable}{lcc}
\tabletypesize{\footnotesize}
\tablecaption{Abundance Ratios at Io \label{tbl-4}}
\tablewidth{0pt}
\tablecolumns{3}
\tablehead{
\colhead{Species} & \colhead{Relative Atmospheric} & \colhead{Reference} \\
\colhead{} & \colhead{Abundance} & \colhead{} \\
}\startdata
Cl/O		&  0.017 $\pm$ 0.008		&  1 \\
Cl/S		&  0.10 $\pm$ 0.05		&  1 \\
S/O		&  0.18 $\pm$ 0.08		&  1 \\
NaCl/SO$_2$	&  7$\times$10$^{-4}$--0.02	&  2 \\
S/SO$_2$	&  (3--7)$\times$10$^{-3}$	&  3 \\
O/SO$_2$	&  0.05				&  4 \\
Cl/SO$_2$	&  (3.0--8.5)$\times$10$^{-4}$	&  1,3,4 \\
\enddata
\tablerefs{
(1) This paper; (2) Lellouch et al. 2003; (3) McGrath et al. 2000; 
(4) Geissler et al. 2004.}
\end{deluxetable}

\clearpage

\begin{figure}
\epsscale{0.95}
\plotone{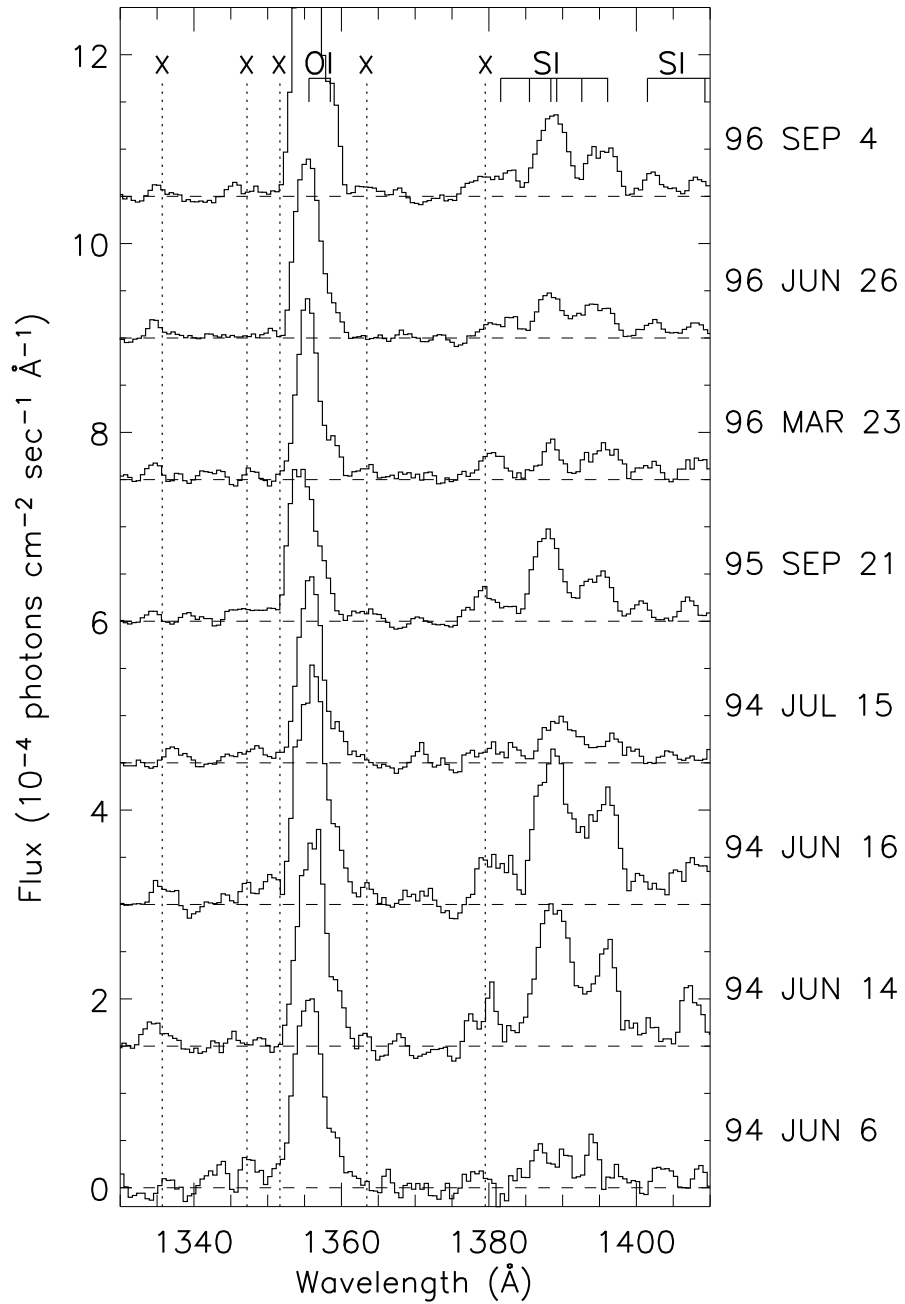}
\caption{Individual flux spectra of Io's atmospheric emissions in the 
wavelength range 1330--1410 \AA\, taken with HST/GHRS from 1994--1996 
to illustrate the variation from date to date and the need for co-adding the 
data to achieve better signal to noise.   All data are on the same 
scale with the zero flux line shifted for display purposes.  The X's indicate 
possible chlorine emission.
\label{fig1}}
\end{figure}

\begin{figure}
\epsscale{0.80}
\rotatebox{90}{
\plotone{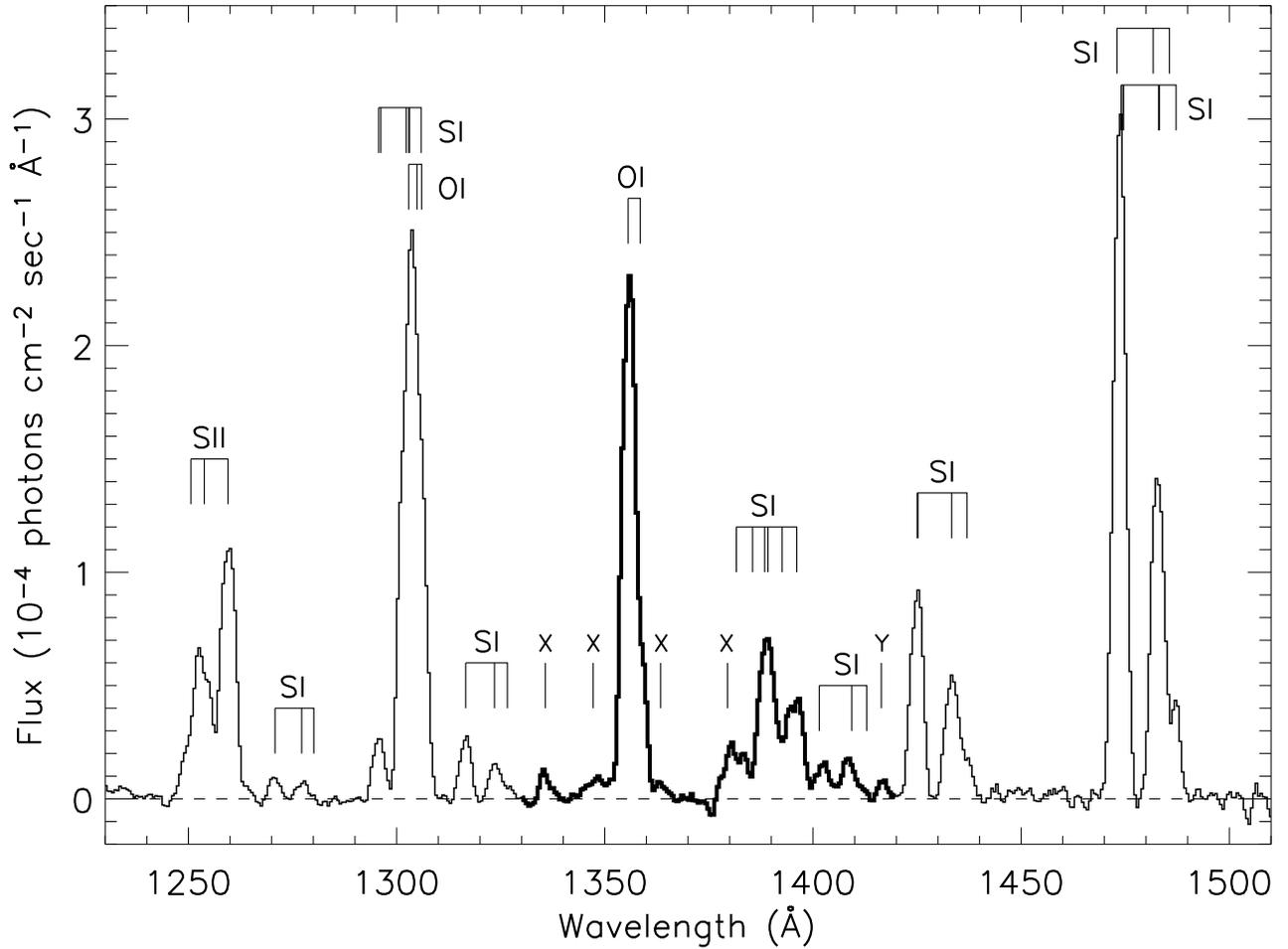}}
\caption{Co-added flux spectrum of Io's atmospheric emissions in the 
wavelength range 1230--1510 \AA\, taken with HST/GHRS from 1994--1996.  
Several sulfur and oxygen features previously identified in Io's atmosphere 
are indicated.  The X's indicate possible chlorine emission.  Y indicates 
possible \ion{S}{4}.  The portion of the spectrum in boldface is analyzed in 
this paper.
\label{fig2}}
\end{figure}

\begin{figure}
\epsscale{0.65}
\plotone{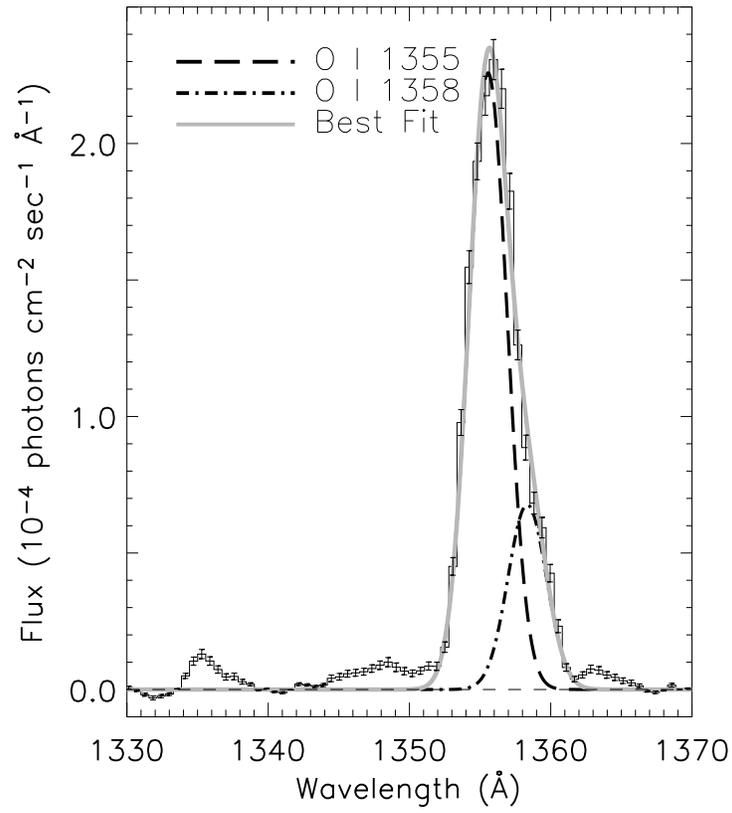}
\caption{Gaussian fits to the \ion{O}{1}] $\lambda$1356 emission in 
the flux spectrum of the co-added GHRS data previously shown in Figure 2.  
The 1355.6 \AA\, (dashed) and 1358.5 \AA\, (dash-dotted) fits are 
shown separately and the total fit is in gray.
\label{fig3}}
\end{figure}

\begin{figure}
\epsscale{0.94}
\plotone{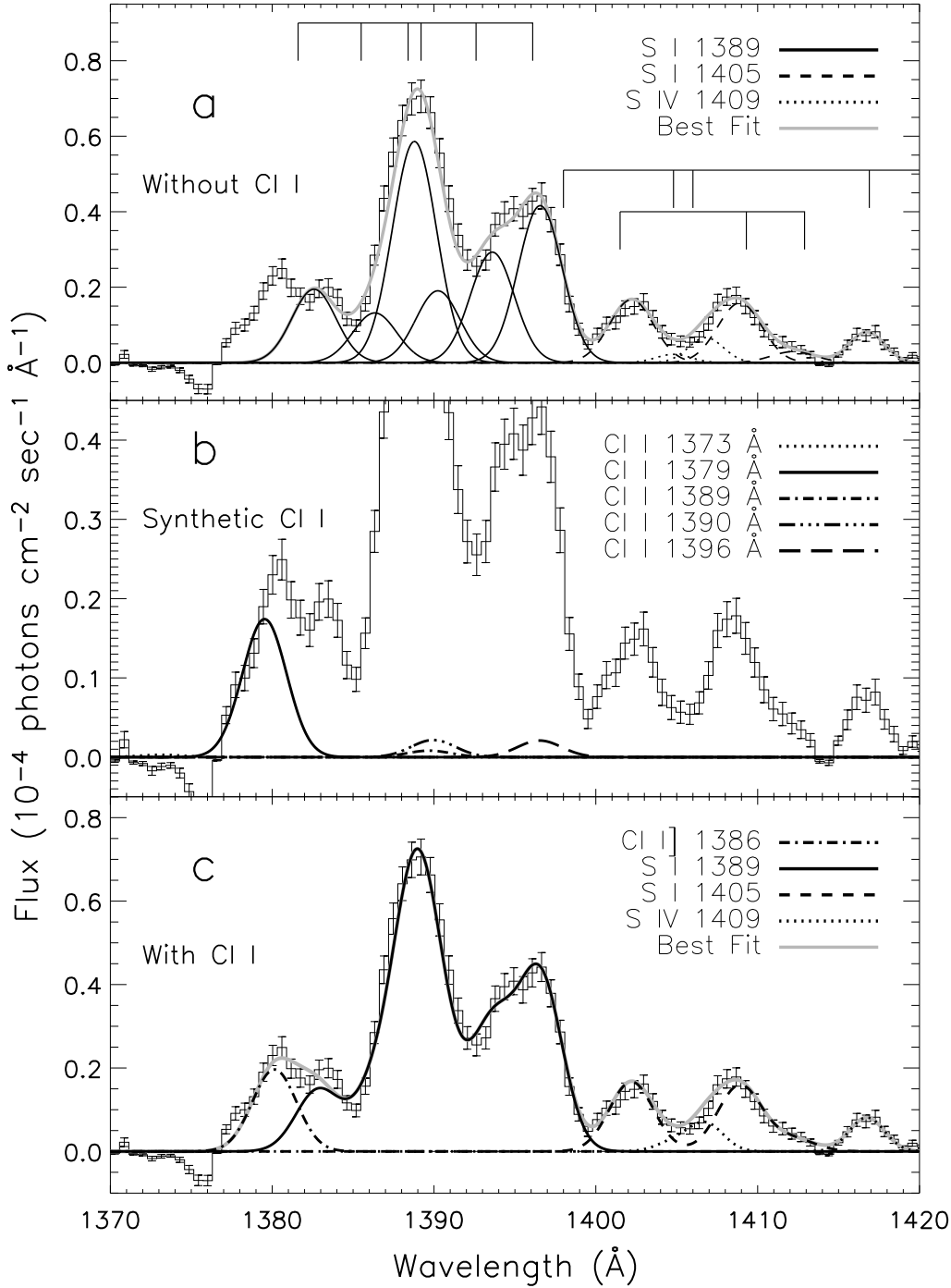}
\caption{Flux spectrum of the GHRS data previously shown in Figure 2, 
limited to the  range 1370--1420 \AA\.  (a)  Gaussian fits to 
\ion{S}{1} $\lambda$1389 (solid), \ion{S}{1} $\lambda$1405 (dashed), and 
\ion{S}{4}] $\lambda$1406 (dotted).  Best total fit is in gray.  Excess flux 
shortward of the \ion{S}{1} $\lambda$1389 multiplet is seen at a SNR of 
$\sim\,$10. 
(b)  Synthetic spectrum of the \ion{Cl}{1}] $\lambda$1386 multiplet assuming 
optically thin line ratios.
(c)  Same as (a) with the Gaussian fit to the detected 1379.5\,\AA\, 
transition of \ion{Cl}{1}] $\lambda$1386 (dash-dotted).  The solid gray line 
is the total fit for the region.
\label{fig4}}
\end{figure}

\begin{figure}
\epsscale{0.89}
\plotone{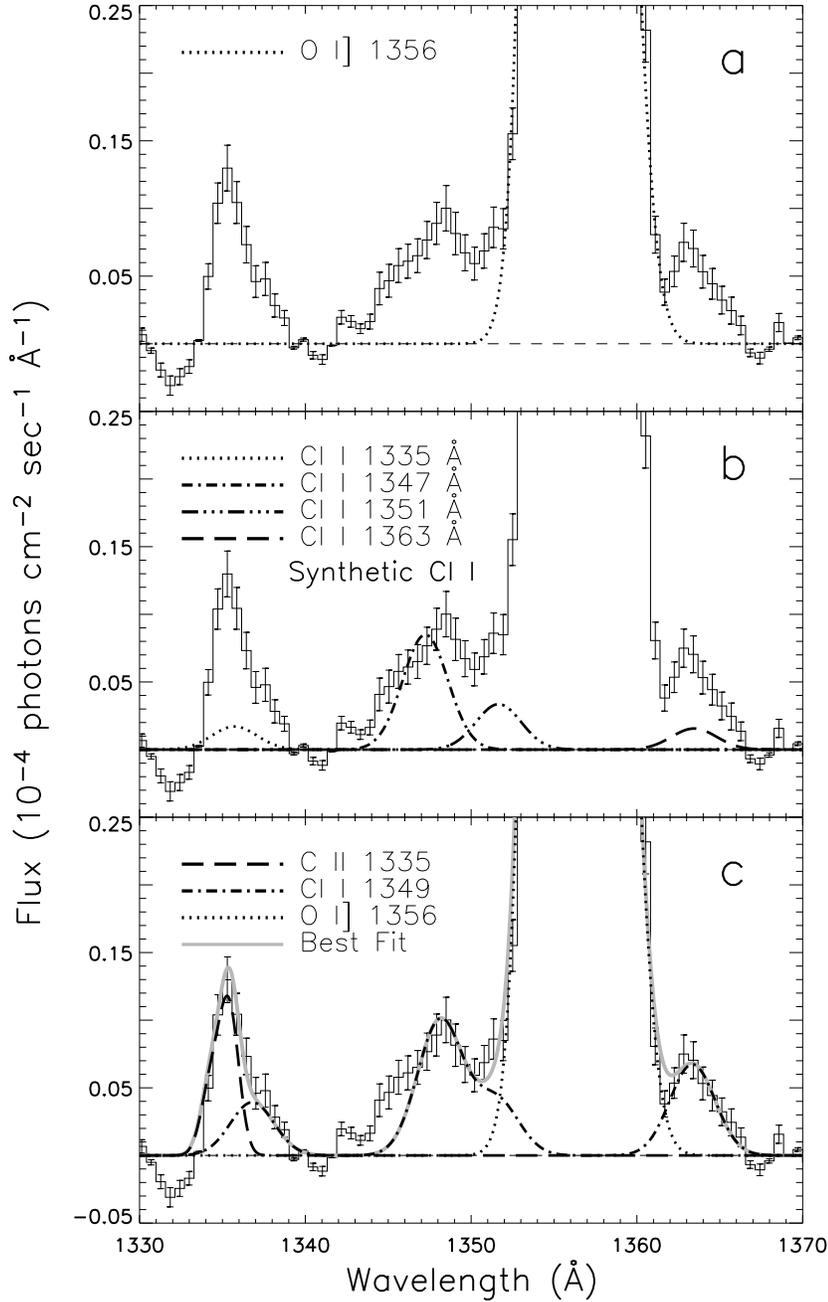}
\caption{(a)  Gaussian fit to the \ion{O}{1}] $\lambda$1356 doublet (dotted) 
previously shown in Figure 3 with excess flux flanking the oxygen emission 
above the noise level.  
(b)  Synthetic spectrum of the \ion{Cl}{1} $\lambda$1349 multiplet emission 
assuming optically thin line ratios.  
(c)  Gaussian fits to the detected \ion{Cl}{1} $\lambda$1349 transitions 
(dash-dotted) are added to the oxygen doublet (dotted).  The chlorine 
transition at 1335 \AA\, is contaminated by a reflected solar component, 
\ion{C}{2} $\lambda$1335, whose fit is shown with a dashed line.  In our best 
fits, the carbon is assumed to account for most of the flux at 
1335\,\AA\, based on the FWHM and amplitude analysis of the line.  See text 
for discussion.  The total fit is in gray.
\label{fig5}}
\end{figure}

\end{document}